\documentclass[10pt]{article}
\usepackage{amsmath}
\usepackage{amssymb}
\usepackage{cite}
\usepackage{lineno}
\usepackage{microtype}
\DisableLigatures[f]{encoding = *, family = * }
\usepackage[labelfont=bf,labelsep=period,justification=raggedright]{caption}
\makeatletter
\renewcommand{\@biblabel}[1]{\quad#1.}
\makeatother
\date{}
\usepackage{graphicx}
\graphicspath{{}}
\DeclareGraphicsExtensions{.pdf,.jpeg,.png,.jpg}

\usepackage{float}

\newcommand{\trials}{250}
\usepackage{grffile}

\newcommand{\figActiveSessions}[0]{
\begin{figure}[H]
    \centering
    \includegraphics[width=\textwidth]{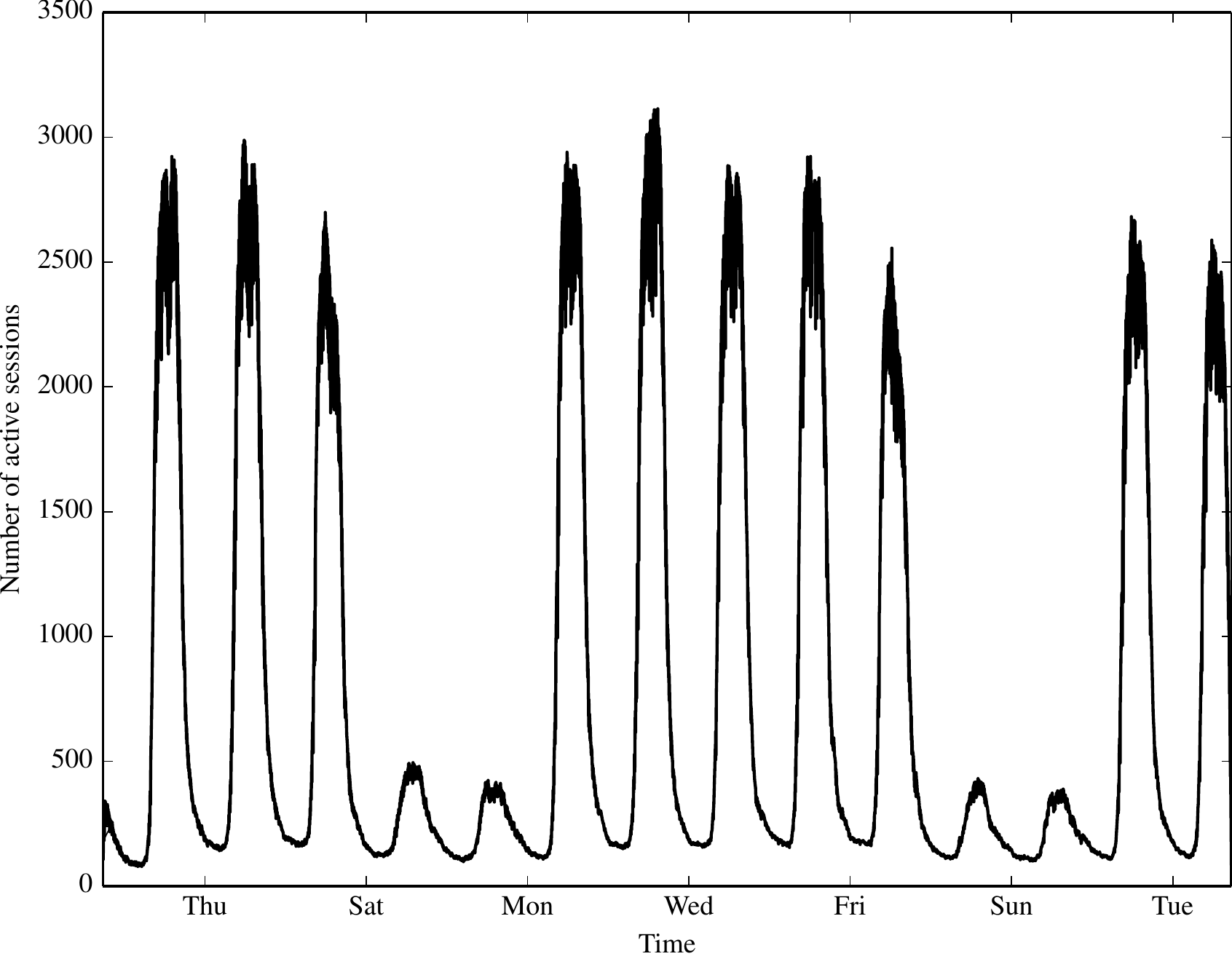}
    \caption{\bf Session volume over time shows diurnal and
        weekday/weekend patterns.}
    \label{fig:active_sessions}
\end{figure}
}
\newcommand{\figLocationsPerNodeECDFLCCTrue}[0]{
\begin{figure}[H]
    \centering
    \includegraphics[width=\textwidth]{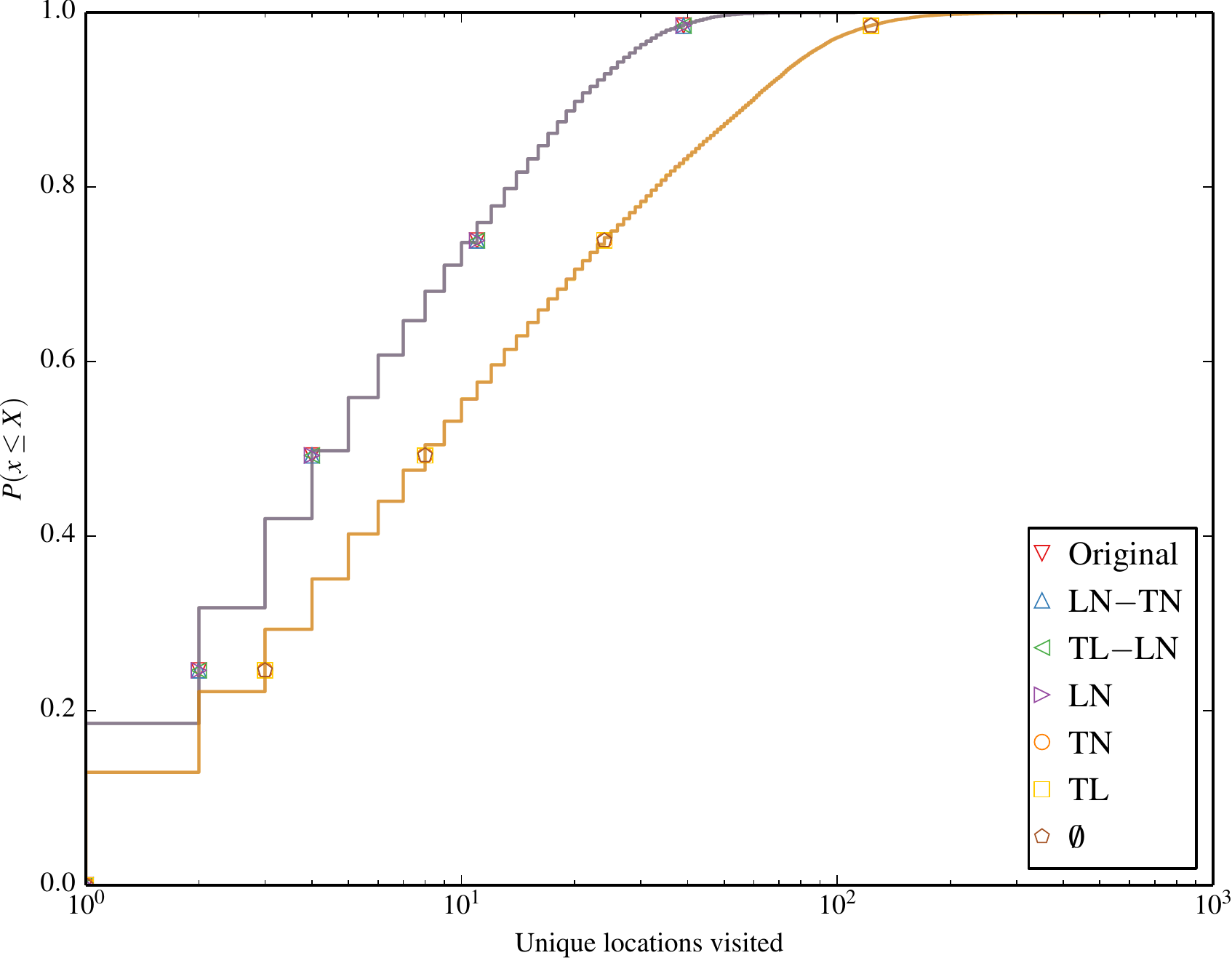}
    \caption{{\bf (color online) Unique locations per node.}  The number of
      locations per node show the increase of unique locations per node when LN
      is broken.}
    \label{fig:locations_per_node_ecdf_lcc_true}
\end{figure}
}
\newcommand{\figContactsPerNodeECDFTotalLCCTrue}[0]{
\begin{figure}[H]
    \centering
    \includegraphics[width=\textwidth]{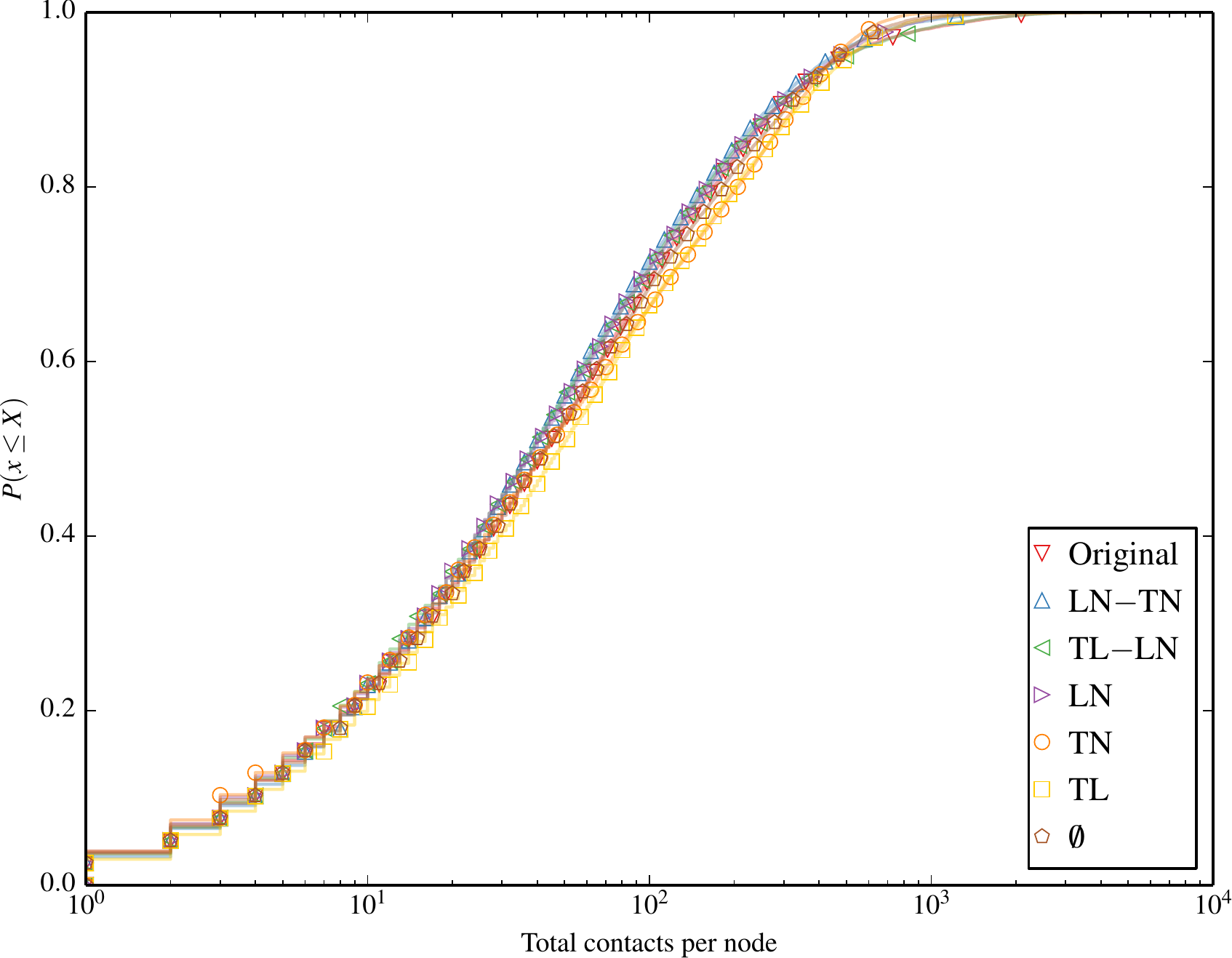}
    \caption{{\bf (color online) Total contacts per node.}
      Total contacts per node are comparable across null models.}
    \label{fig:contacts_per_node_ecdf_total_lcc_true}
\end{figure}
}
\newcommand{\figContactsPerNodeECDFUniqLCCTrue}[0]{
\begin{figure}[H]
    \centering
    \includegraphics[width=\textwidth]{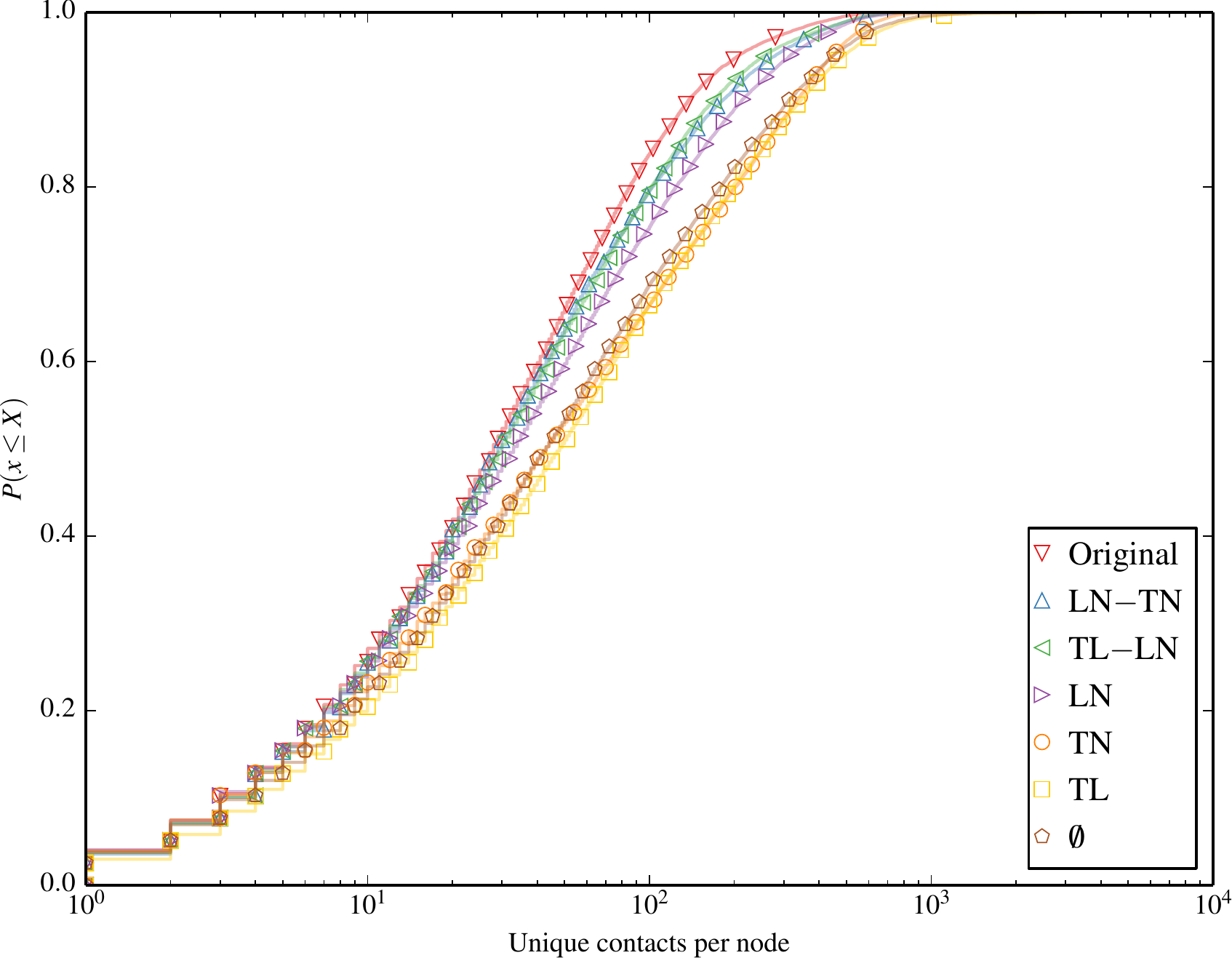}
    \caption{{\bf (color online) Unique contacts per node.}
      Null models that destroy LN result in higher unique contacts.}
    \label{fig:contacts_per_node_ecdf_uniq_lcc_true}
\end{figure}
}
\newcommand{\figIntersessionTimeECDFLCCTrue}[0]{
\begin{figure}[H]
    \centering
    \includegraphics[width=\textwidth]{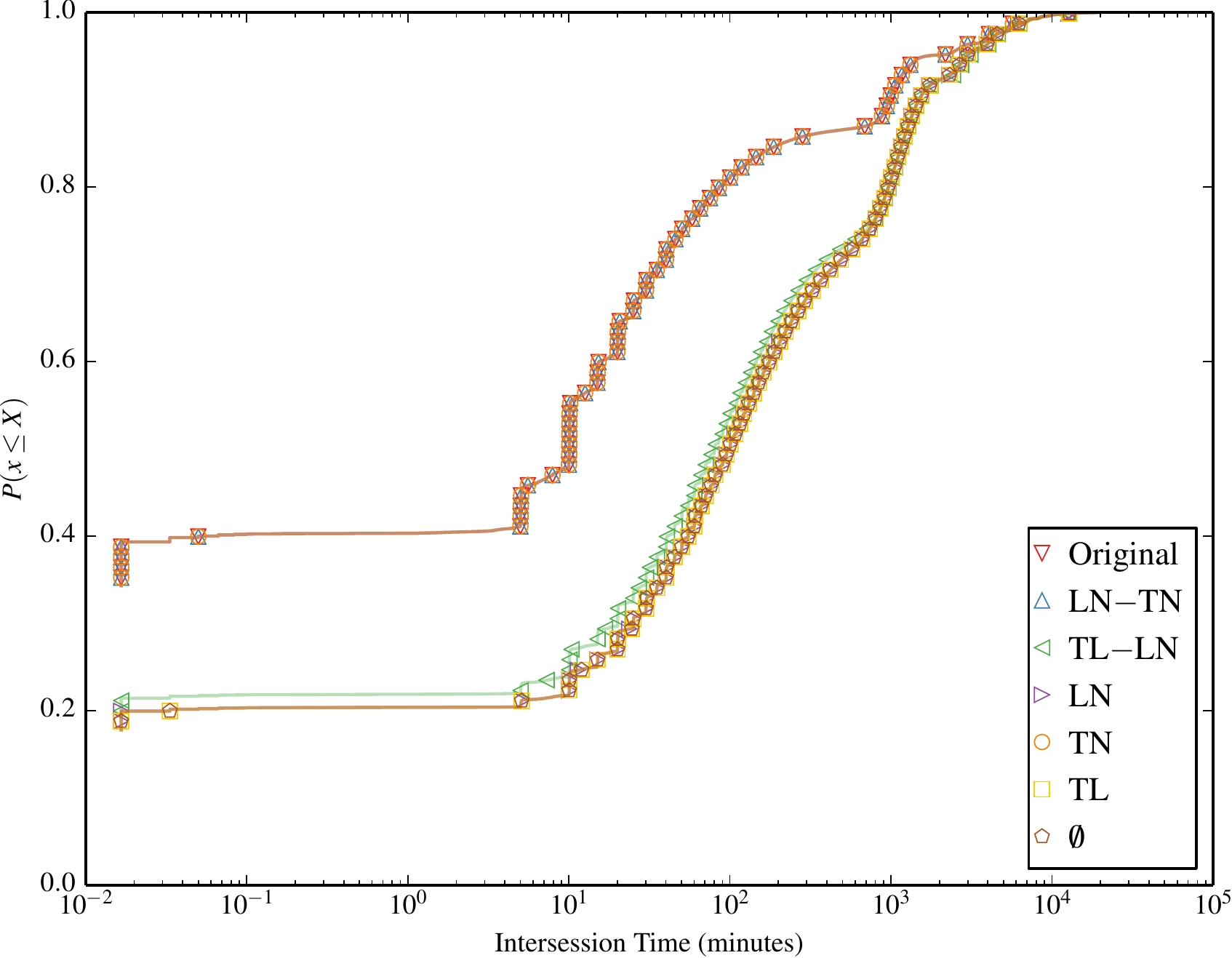}
    \caption{{\bf (color online) Intersession time.}  The intersession time
      distribution for each node is clearly longer for all null models that
      break TN.}
    \label{fig:intersession_time_ecdf_lcc_true}
\end{figure}
}
\newcommand{\figSBSWPrevVsTimeLCCTrue}[0]{
\begin{figure}[H]
    \centering
    \includegraphics[width=\textwidth]{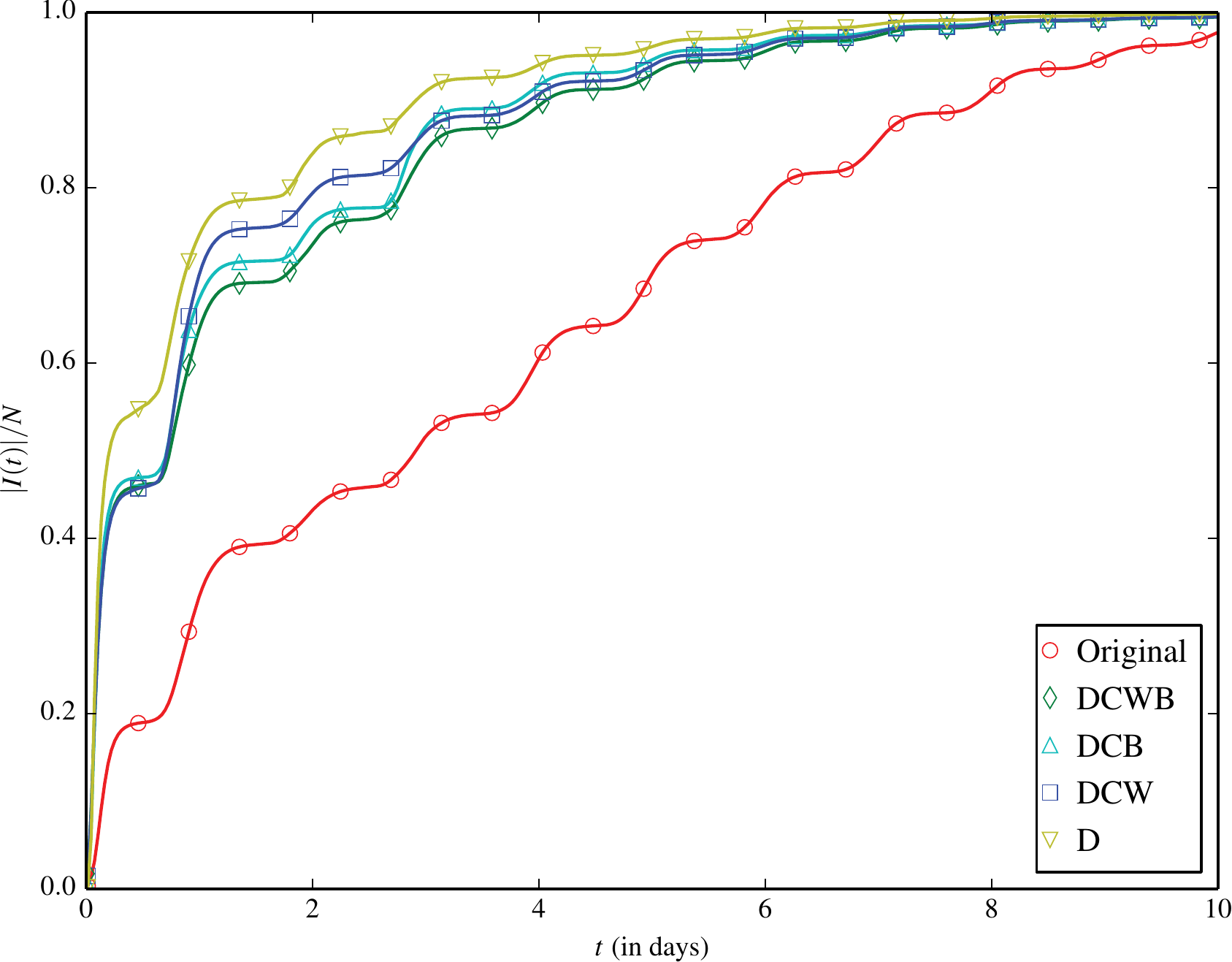}
    \caption{{\bf (color online) Fraction of infected devices $|I(t) |/N$ as a
        function of time, for the original contact trace and ``Small But Slow
        World'' null model contact shufflings.}  Each line is based on a
        uniformly spaced sampling of prevalence over time.  Note that the
        sampling resolution of lines is higher than their respective markers, as
        we only plot every $n$-th marker ($n > 1$) to minimize visual clutter.
        This (not fitting) is the reason for nonlinearity between plotted
        points.  The same applies for the other plots presented throughout this
        paper.  }
    \label{fig:sbsw_prev_vs_time_lcc_true}
\end{figure}
}
\newcommand{\figSSRContactCountVsTimeTotal}[0]{
\begin{figure}[H]
    \centering
    \includegraphics[width=0.7\textwidth]{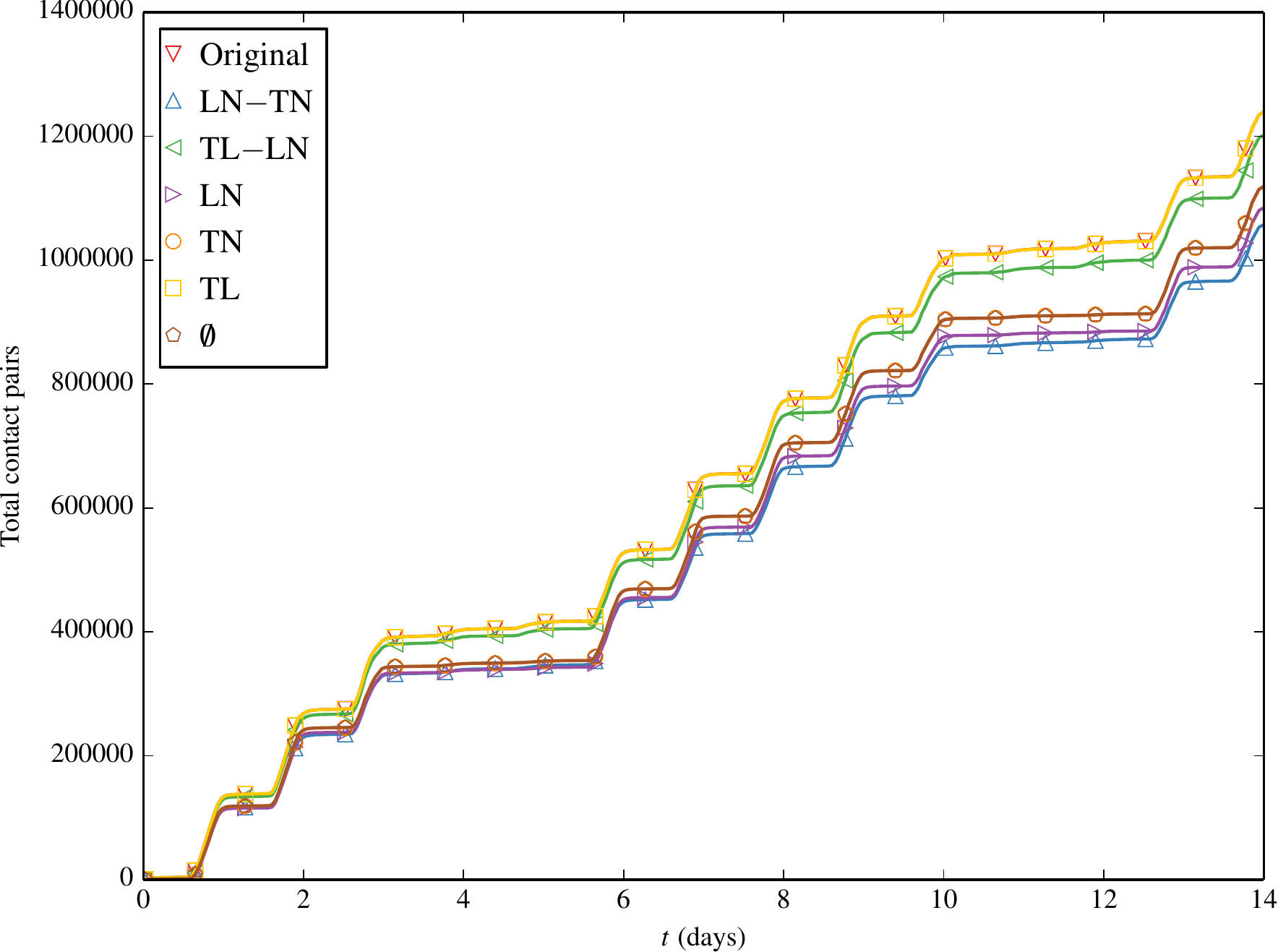}
    \caption{\bf{(color online) Total cumulative contacts as a function of time,
        for the original contact trace and contact traces induced by inducement
        shuffling.}}
    \label{fig:ssr_contact_count_vs_time_total}
\end{figure}
}
\newcommand{\figSSRContactCountVsTimeUnique}[0]{
\begin{figure}[H]
    \centering
    \includegraphics[width=0.7\textwidth]{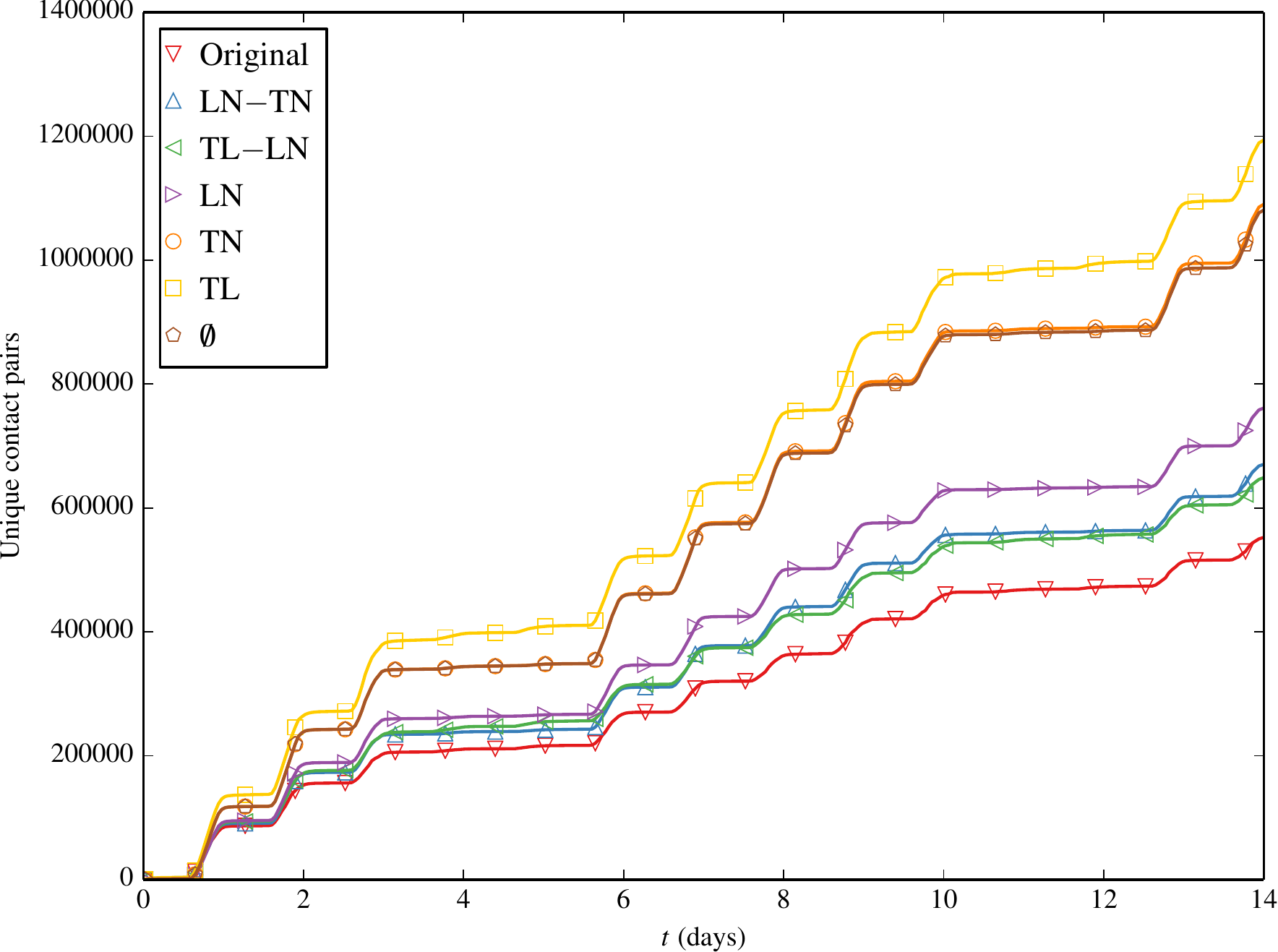}
    \caption{\bf{(color online) Unique cumulative contacts as a function of
        time, for the original contact trace and contact traces induced by
        inducement shuffling.}}
    \label{fig:ssr_contact_count_vs_time_unique}
\end{figure}
}
\newcommand{\figSSRPrevVsTimeLCCTrue}[0]{
\begin{figure}[H]
    \centering
    \includegraphics[width=\textwidth]{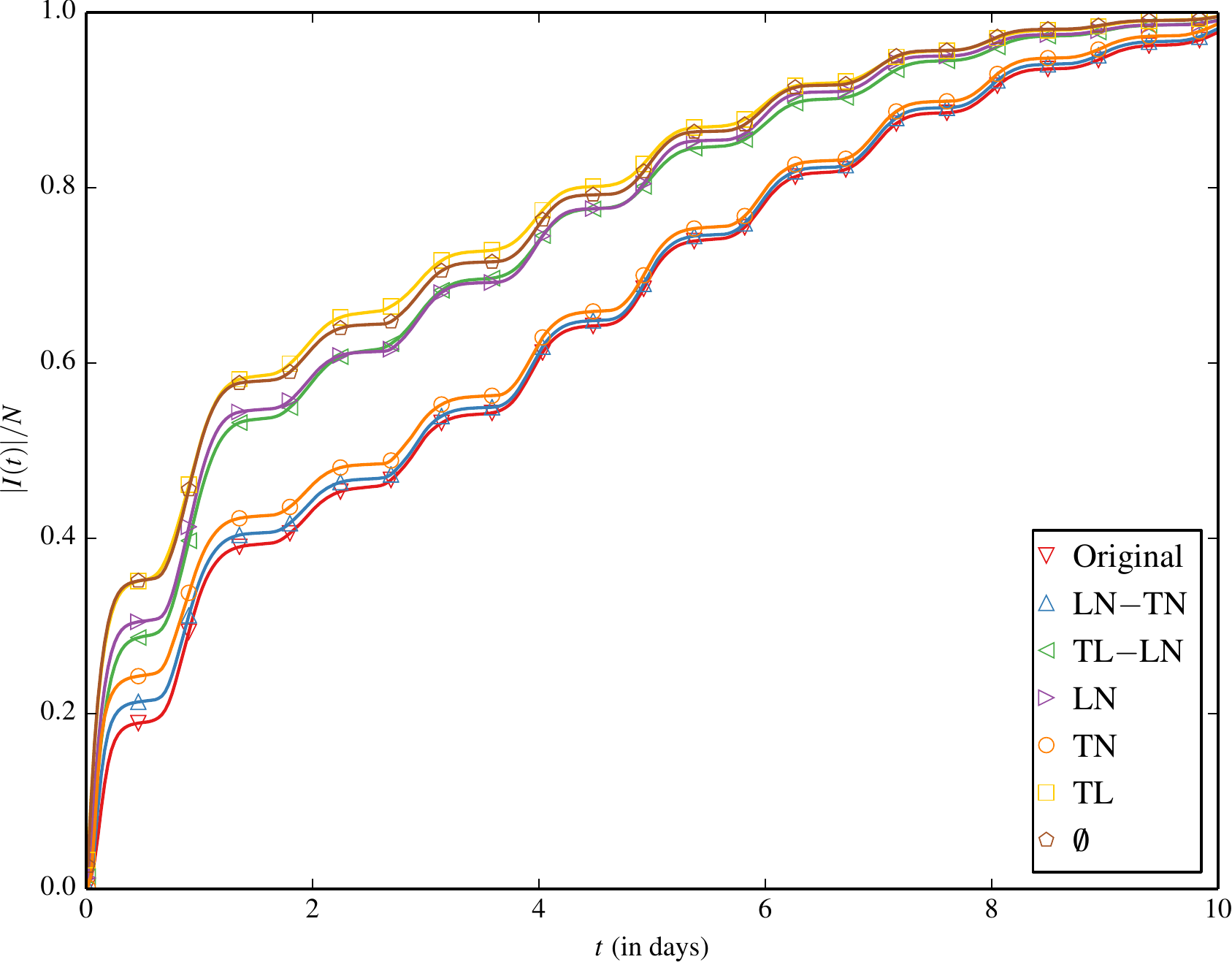}
    \caption{{\bf(color online) Fraction of infected devices $|I(t)|/N$ as a
        function of time, for the original contact trace and inducement-shuffled
        null models.}}
    \label{fig:ssr_prev_vs_time_lcc_true}
\end{figure}
}
\newcommand{\figOneDayPrevsHists}[0]{
\begin{figure}[H]
    \centering
    \includegraphics[width=\textwidth]{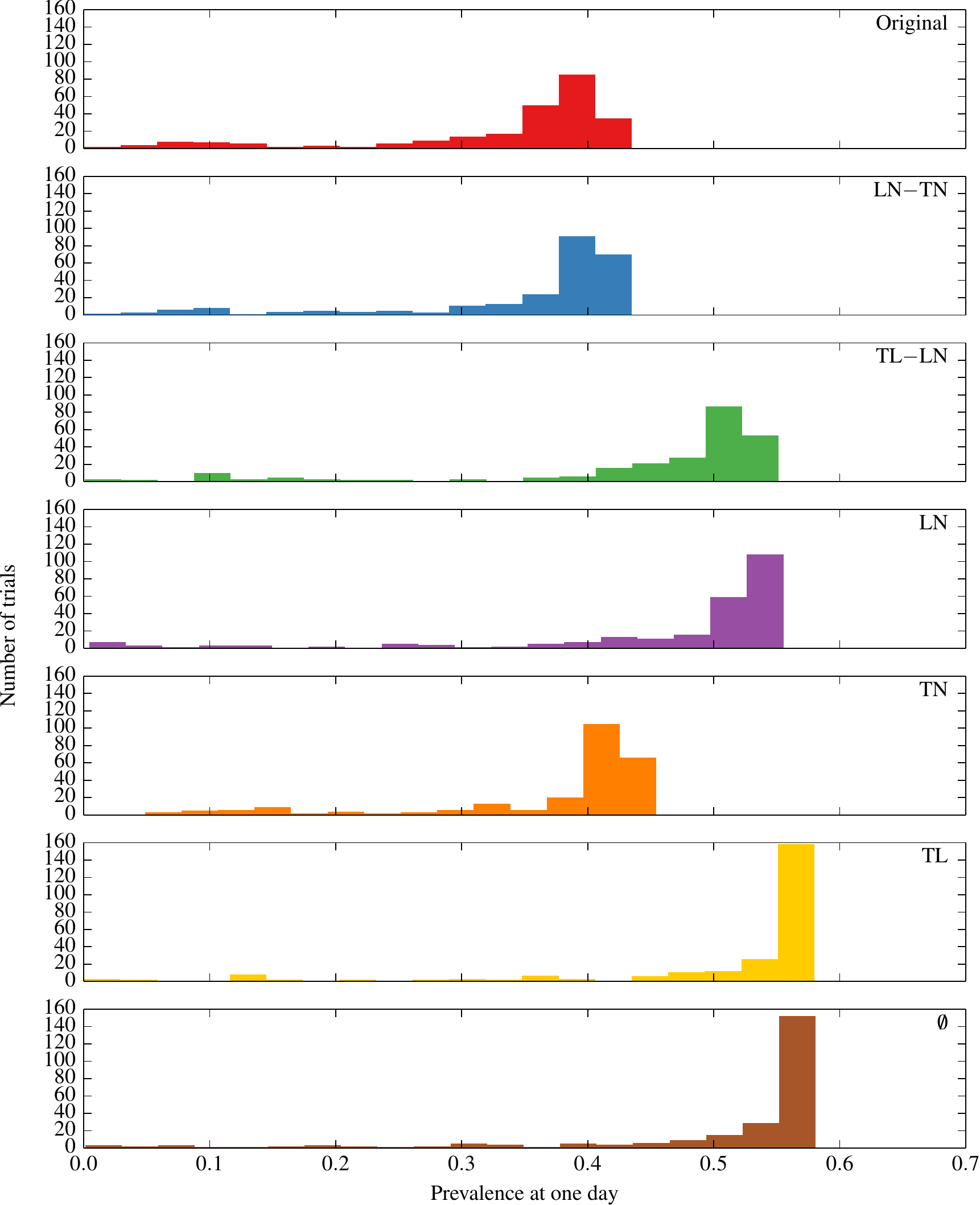}
    \caption{{\bf(color online) The histograms of the prevalence after one day under each null model. }}
    \label{fig:one_day_prevs_hists}
\end{figure}
}
\newcommand{\figSSRPrevVsContactsTotalLCCTrue}[0]{
\begin{figure}[H]
    \centering
    \includegraphics[width=\textwidth]{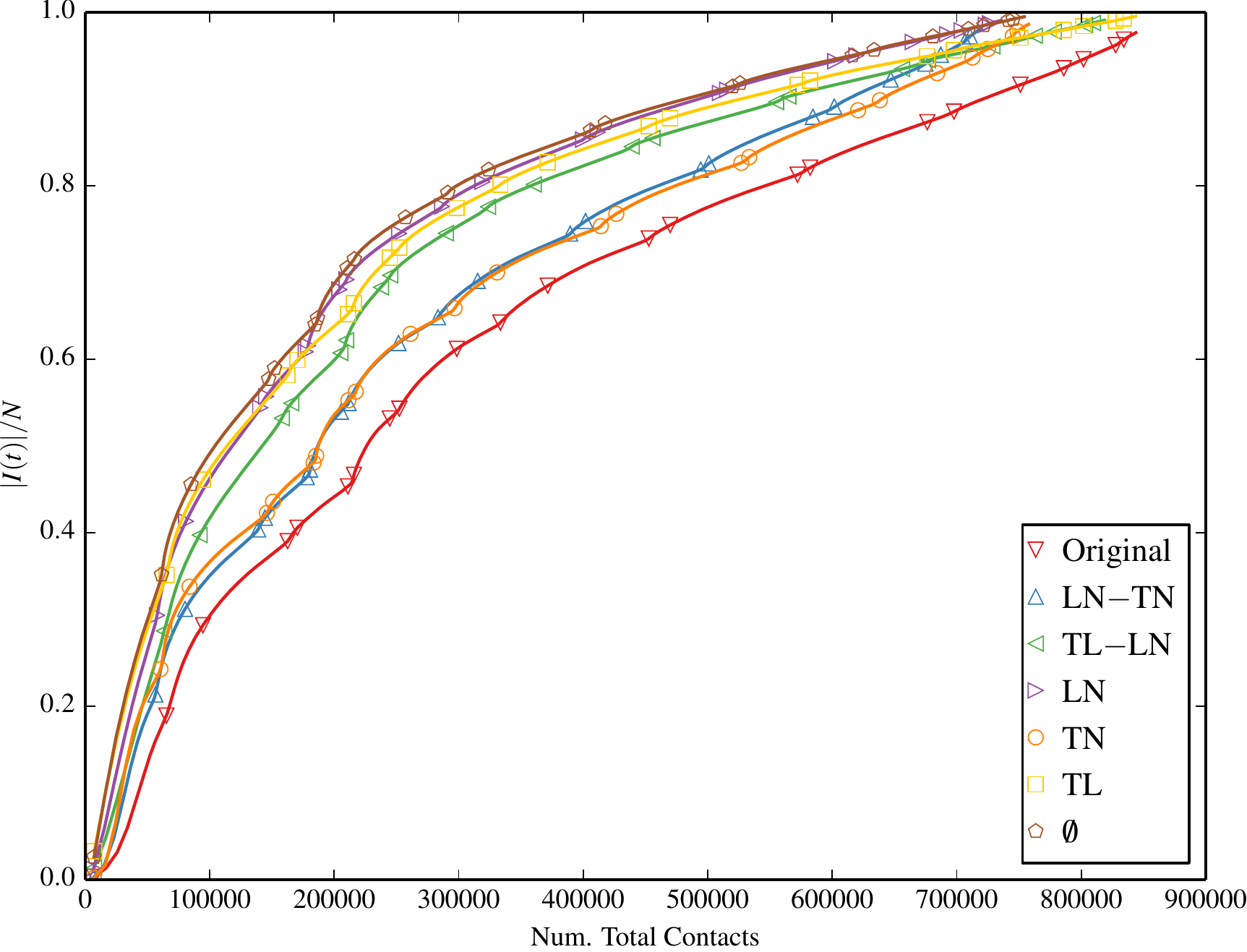}
    \caption{{\bf(color online) Fraction of infected devices $|I(t)|/N$ as a
        function of the number of \emph{total} contacts, for the original
        contact trace and inducement-shuffled null models.}}
    \label{fig:ssr_prev_vs_contacts_total_lcc_true}
\end{figure}
}
\newcommand{\figSSRPrevVsContactsUniqueLCCTrue}[0]{
\begin{figure}[H]
    \centering
    \includegraphics[width=\textwidth]{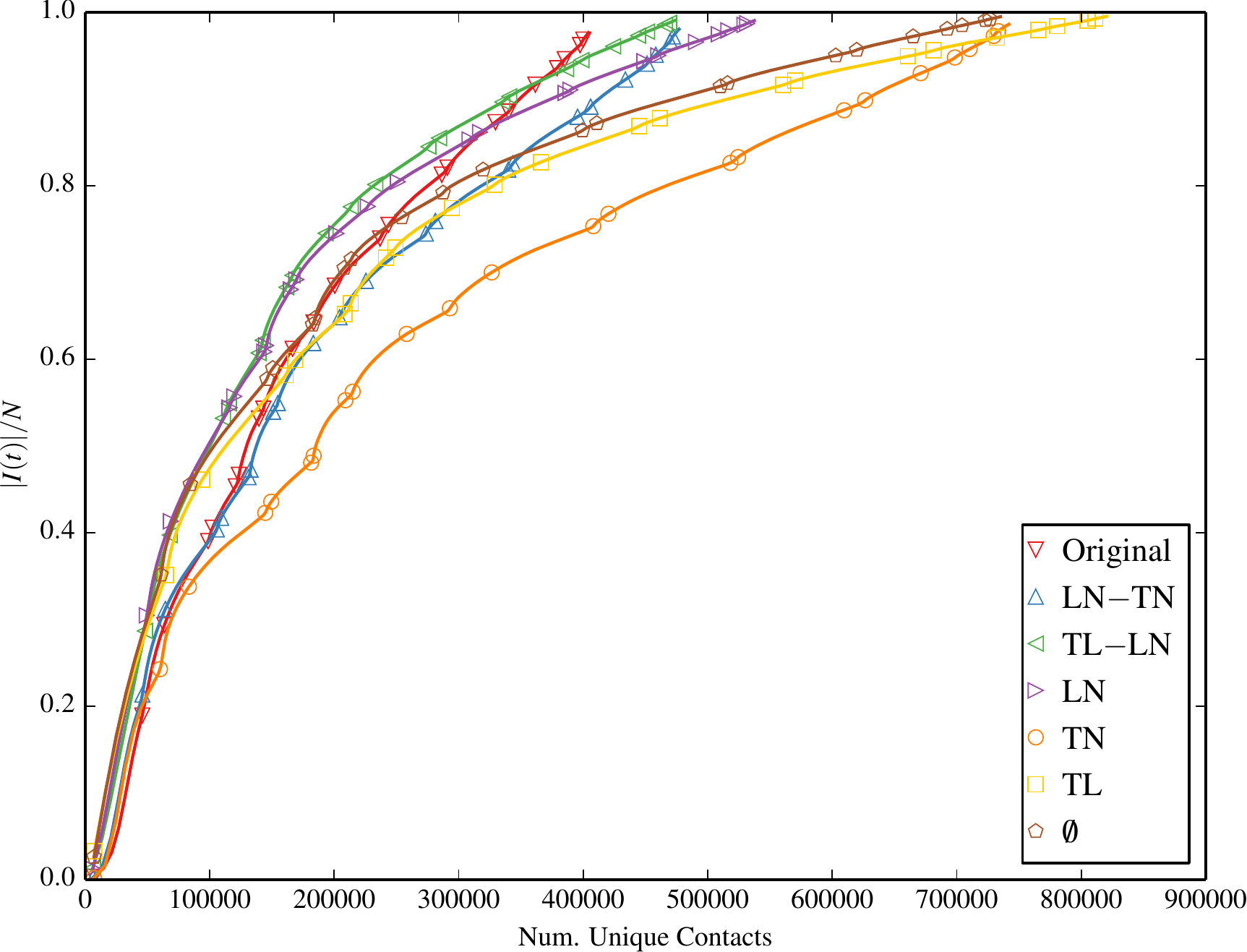}
    \caption{{\bf(color online) Fraction of infected devices $|I(t)|/N$ as a
        function of the number of \emph{unique} contacts, for the original
        contact trace and inducement-shuffled null models.}}
    \label{fig:ssr_prev_vs_contacts_unique_lcc_true}
\end{figure}
}
\newcommand{\figSSRPrevDeltaSEMPairsLCCTrue}[0]{
\begin{figure}[H]
    \centering
    \includegraphics[width=0.9\textwidth]{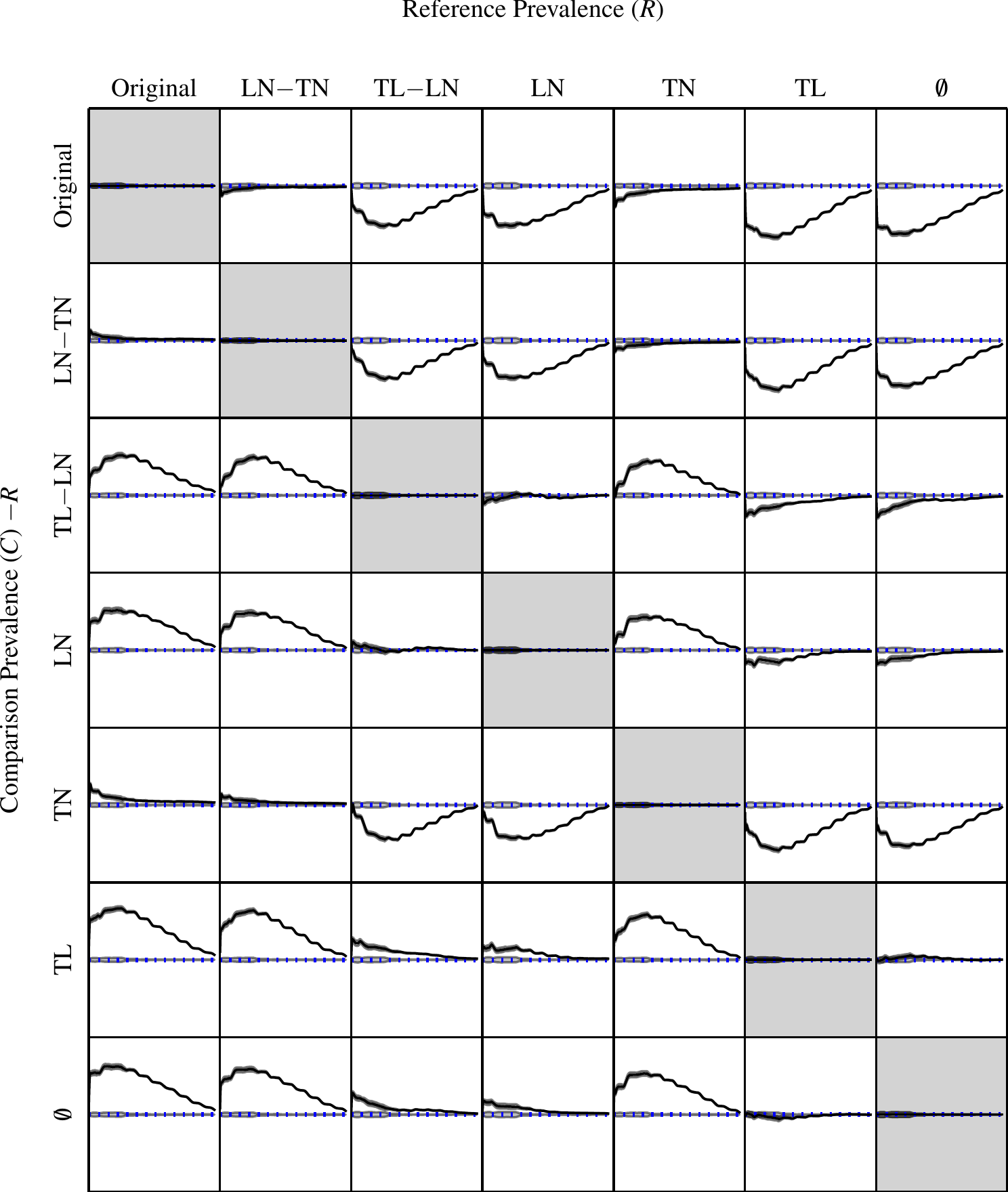}
    \caption{{\bf(color online) Macroscale prevalences over time of all shuffling pairs
        compared.}  The x-axis of each subplot is time and spans the 10-day
        simulation interval, domain [0,10].  The y-axis of each subplot is the
        column label's prevalence subtracted from the row label's prevalence and
        spans the range interval [-0.3,0.3] i.e. 30\% either side of the blue
        dotted line which is centered at 0.0.  The gray shaded region around
        each black line (very small in most plots but visible at high zoom) is
        the standard error of the mean of the row label's prevalence.  The gray
        shaded region around the center dotted blue line is the standard error
        of the mean of the column label's prevalence.}
    \label{fig:ssr_prev_delta_sem_pairs_lcc_true}
\end{figure}
}
\newcommand{\figSSRPrevDeltaSEMPairsLCCTrueAbsLim}[0]{
\begin{figure}[H]
    \centering
    \includegraphics[width=0.9\textwidth]{{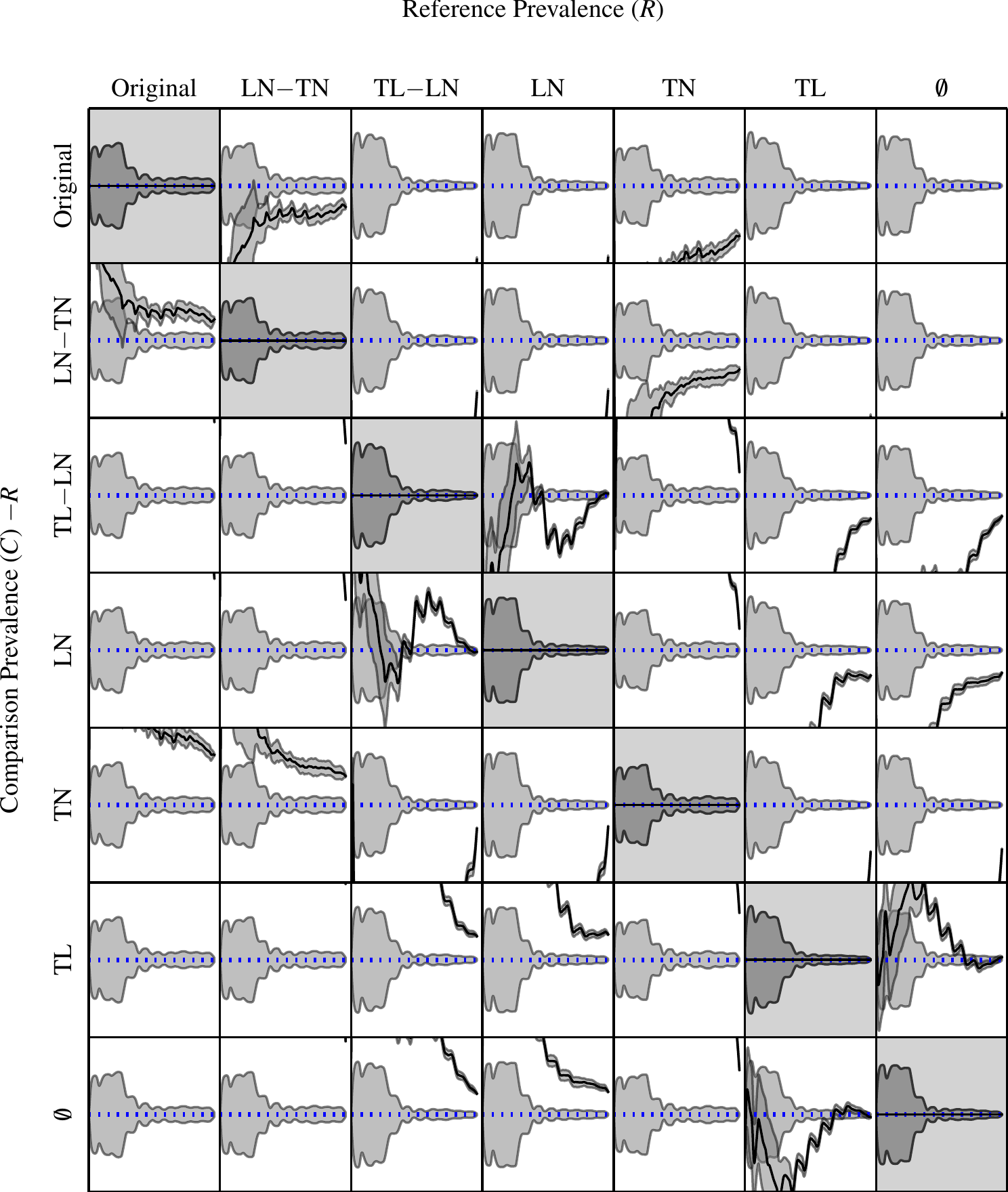}}
    \caption{{\bf(color online) Microscale prevalences over time of all
        shuffling pairs compared.}  This figure is the same as
      Figure~\ref{fig:ssr_prev_delta_sem_pairs_lcc_true}, the only difference
      being that the y-axis range has been zoomed to the interval [-0.015,0.015]
      i.e. 1.5\% either side of the dotted blue line instead of 30\% either
      side.  This highlights the standard error of the mean more clearly and
      allows one to more easily establish regions of overlap between the
      reference and comparison's standard errors giving a better idea of where
      differences are of lower statistical significance.}
    \label{fig:ssr_prev_delta_sem_pairs_lcc_true_abs_lim}
\end{figure}
}

\begin{document}

\begin{flushleft}
{\Large
\textbf{Diffusion in Colocation Contact Networks: the Impact of Nodal Spatiotemporal Dynamics}
}
\\
Bryce Thomas$^{1,\ast}$, Raja Jurdak$^{2}$, Kun Zhao$^{2}$, Ian Atkinson$^{3}$
\\
\bf{1} Department of Information Technology, James Cook University, Townsville,
Queensland, Australia
\\
\bf{2} Autonomous Systems, CSIRO, Brisbane, Queensland, Australia
\\
\bf{3} eResearch, James Cook University, Townsville, Queensland, Australia
\\
$\ast$ E-mail: bryce.m.thomas@gmail.com (BT)
\end{flushleft}

\section*{Abstract}
Temporal contact networks are studied to understand dynamic spreading phenomena
such as communicable diseases or information dissemination. To establish how
spatiotemporal dynamics of nodes impact spreading potential in colocation
contact networks, we propose ``inducement-shuffling'' null models which break
one or more correlations between times, locations and nodes.  By reconfiguring
the time and/or location of each node's presence in the network, these models
induce alternative sets of colocation events giving rise to contact networks
with varying spreading potential.  This enables second-order causal reasoning
about how correlations in nodes' spatiotemporal preferences not only lead to a
given contact network but ultimately influence the network's spreading
potential.  We find the correlation between nodes and times to be the greatest
impediment to spreading, while the correlation between times and locations
slightly catalyzes spreading. Under each of the presented null models we measure
both the number of contacts and infection prevalence as a function of time, with
the surprising finding that the two have no direct causality.

\section*{Introduction}\label{sec:intro}
Complex networks~\cite{statistical_mechanics, newman2003structure,
  barrat2008dynamical} as a unified framework can describe a wide range of
systems including the Internet and World Wide Web~\cite{scale_free}, biological
systems~\cite{jeong2000large, graph_theory_and_networks_in_biology,
  bullmore2009complex} and social
networks~\cite{redner1998popular,newman2004coauthorship,eagle2009inferring}. Dynamical
processes taking place on networks, such as information diffusion or epidemic
spreading, are strongly influenced by the structure and organization of
networks~\cite{barrat2008dynamical}. For a long time, studies on the interplay
between network structure and dynamical processes have been focused on static
networks~\cite{pastor2001epidemic, newman2002spread}. Recently, attention has
turned to the study of more complicated systems, in which dynamical processes
are taking place on temporal networks with time-varying
structures~\cite{temporal_networks, liu2013contagion}. Underlying factors in
these systems that drive the evolution of temporal networks can subsequently
impact the embedding dynamic phenomenon. Identifying these factors and their
intrinsic association with the process is crucial for unravelling the complexity
of these systems, as well as modelling, predicting and controlling the dynamic
phenomenon.

Contact networks are one important instantiation of temporal networks that model
either the logical or physical contacts between individuals.  Examples of the
former include long-range communications such as phone calls or email.  Examples
of the latter include colocation-driven short-range communications such as
face-to-face human contact~\cite{zhao2011social, starnini2013modeling} or
proximity-based direct wireless transmissions~\cite{trace_based_mobility}.
Unlike logical contact networks, physical (i.e. colocation) contact networks are
predicated on spatially constrained copresence and depend on mobility for
broadscale spreading.  The spatiotemporal dynamics of the actors (nodes) in such
networks are therefore a critical determinant of spreading potential.

One of the best available sources of empirical data on colocation contact (also
known as ``encounter'') networks comes from electronic mobile wireless device
traces~\cite{trace_based_mobility}.  These traces typically describe either
directly recorded encounters between devices~\cite{reality_mining,
  pocket_switched, user_mobility_for_opportunistic, ONE} or encounters which are
inferred based on mutual presence at a known location~\cite{ONE,
  content_diffusion_in}.  Much work has been devoted to the analysis of these
traces, often in terms of their static network-theoretic properties and in some
instances also in terms of spreading potential.  For example, the authors
in~\cite{ONE, user_mobility_for_opportunistic, content_diffusion_in} all
explicitly analyzed ad hoc~\cite{routing_in_ad_hoc} multi-hop message
dissemination~\cite{delay_tolerant_networking} facilitated by device mobility
and encounters, the latter work being our own performed over the same trace
analyzed in this paper.

One of the core tools for relating observed spreading behavior of a network
quantity (e.g. virus, information) to idiosyncratic network features are
\emph{null models}~\cite{SBSW, gauvin2013activity} which separately destroy
correlations in a network's contact structure.  By simulating spreading on both
the real and null networks one can establish to what extent certain correlations
catalyze or impede spreading.  Prior null models~\cite{SBSW} have focused on
\emph{contact} shuffling---given a set of nodes connected by timestamped links,
these models shuffle the links in one of a number of ways to retain certain
correlations while destroying others.  For example, the authors in~\cite{SBSW}
applied null model contact shufflings to a mobile call network (MCN) in which
nodes were mobile subscribers and links were placed between caller/callee pairs
and annotated with one or more call timestamps.  A limitation of contact
shuffled null models as applied to colocation-based contact networks is that
they cannot draw any association between spatiotemporal dynamics and spreading.
Rather, the models consider the contact network independent of the nature of the
events which led to the contacts in the first place, such as the movement
patterns of individual nodes in colocation contact networks.

While many attempts have been made to model such networks, there is yet no
structured framework for establishing how nodes' spatiotemporal preferences
impede or catalyze spreading potential. This paper focuses on exploring the
relationship between node spatiotemporal preferences and spreading potential in
colocation contact networks.  To address this question, we propose an
alternative way of generating network null models through a method we refer to
as \emph{inducement} shuffling---shuffle the events on which contacts are
themselves predicated and so in the process induce a modified set of contact
events.  Inducement shuffling possesses the attractive property of enabling
second-order causal reasoning about spreading behavior.  With contact shuffling
it is only possible to state how characteristics of the contact network itself
influence spreading.  With inducement shuffling one can state how the nature of
the events that lead to the contact network in the first place, such as the node
movement dynamics, influence spreading. While both shuffling approaches lead to
changes in network structure, inducement shuffling explicitly captures the
relationships between the node preferences and correlations on one hand, and the
spreading potential on the other.

We present our inducement-shuffled null models based on the theme of node,
location and time which we use to decorrelate spatiotemporal node preferences.
We apply the null models to an empirical trace in which (i) nodes are mobile
wireless electronic devices such as smartphones, tablets and laptop computers
(ii) locations are Access Points (APs) in a large university campus network and
(iii) times are the times at which given devices were connected to given APs.
Contacts in our empirical trace are predicated on device colocation.  That is,
two devices simultaneously present at a given location are connected by a
timestamped link in the inferred contact network.  Through inducement shuffling
we destroy node's time and location preferences, i.e. we decorrelate the
relationships between nodes, times and locations.  This in turn leads to a
different set of colocation events and thus a modified contact network is
induced.  For the original and each induced contact network we simulate
diffusion of a quantity starting from a randomly infected node under the
Susceptible-Infected (SI) infection model~\cite{infectious_diseases}.  To the
best of the authors' knowledge, the present work is the first to take a null
models approach to isolating spreading impediments and catalysts in
colocation-driven contact networks.  More importantly, we believe the
inducement-shuffled null models to be the first to enable reasoning about the
second-order causal relationship between the events on which the contact network
is predicated and subsequent spreading potential.  Though motivated in the
context of a wireless mobile device contact trace, we believe the
inducement-shuffled null models presented in this paper may find broader
applications in colocation contact networks, some well outside computer
networks.

\section*{Materials and Methods}
\subsection*{Dataset}
Our work utilizes a wireless IEEE 802.11 (Wi-Fi) network trace collected by the
University of Queensland (UQ) Australia describing device sessions at individual
access points (APs) in a large university network over a 14 day period between
Tue Nov 27 17:39:12 AEST 2012\unskip--%
\makeatletter{}%
Tue Dec 11 17:29:16 AEST 2012
.
Each session record in the UQ trace includes (i) the unique MAC address of the
connecting mobile device (ii) session start time (iii) session end time (iv) AP
name and (v) site.  We clean the trace by discarding 1,462 sessions with no end
time (active when trace collection ceased) and 1 session with zero duration
(start $=$ end).  A further 1,279 sessions are discarded as specifying no AP.
After cleaning, the UQ trace retains %
\makeatletter{}%
546,260 %
sessions
from %
\makeatletter{}%
23,895 %
devices over
\makeatletter{}%
3,079 %
APs at a total of 24 discrete geographic sites.
In this paper we focus only on the largest site in the UQ trace---the \emph{St
  Lucia} campus.  The St Lucia campus is a large university campus accounting
for %
\makeatletter{}%
445,867 %
sessions from
\makeatletter{}%
20,308 %
devices over
\makeatletter{}%
2,004 %
APs.  After filtering on (v) to extract the
St Lucia trace from the UQ trace, we then use (i)--(iv) to construct the minimum
dataset for our analysis which consists of a set of session 4-tuples, each of
the form $\langle N, T_{start}, T_{end}, L\rangle$.
These 4-tuples fully describe which ($N$)ode (MAC address) partakes in a session
at what ($T$)ime (start and end) and at what ($L$)ocation (AP), i.e. they encode
the information about the spatiotemporal patterns of nodes.  We refer to these
4-tuples simply as the ``session tuples'' from this point onwards.  Note that
nodes and locations just happen to be defined in terms of MAC addresses and APs
respectively in our analysis.  The notion of nodes and locations may be
generalized to accommodate equivalent entities specific to the network under
consideration.

\subsection*{Contact Inference}
The session tuples extracted from our original dataset and later reconfigured by
our inducement shufflings do not explicitly describe contacts between pairs of
wireless mobile devices.  Rather, as is done in~\cite{ONE} we infer contact from
colocation---two devices connected to the same AP at the same time are assumed
to be in transmission range and so inferred as making ``contact''.  As noted by
the authors in ~\cite{ONE}, this inference is an approximation.  Devices
connected to the same AP may not be able to communicate directly, devices
connected to different APs may be able to communicate and some contacts take
place outside the range of APs.  Still, it is believed to be a reasonable
approximation in the network under consideration.  Contact inference translates
the session tuples into a set of 5-tuples of the form
$\langle N_{i}, N_{j}, T_{start}, T_{end}, L \rangle$, describing the location
and duration of a contact between devices $N_{i}$ and $N_{j}$ ($i \ne j$).  For
simplicity, and without loss of generality, we rely only on the first three
fields of the 5-tuples in this paper i.e.
$\langle N_{i}, N_{j}, T_{start} \rangle$, which describe the initiation of
contact events.  We refer to these 3-tuples simply as the ``contact tuples''
from this point onwards. The contact duration is not considered here due to the
lack of explicit session termination in the dataset, which creates a degree of
uncertainty around the exact timing of session completion. Though this
uncertainty is limited to the order of minutes, the incorporation of this
uncertainty into the shuffling approach is out of scope of this paper and is
left for future work.

\subsection*{Contact Shuffling}
As a preamble to our main inducement shuffling results, we apply the
pre-existing contact-shuffled null models presented by the authors in ``Small
but slow world''~\cite{SBSW} to our originally inferred contact network.  This
allows us to compare and contrast spreading behavior against that observed in
prior work.  We refer to these earlier null models simply as the ``SBSW'' or
``contact-shuffled'' null models.  The SBSW models were previously applied to a
number of contact networks, most notably a large Mobile Call Network (MCN) with
timestamped links between caller/callee pairs.  The authors use strings of
capital letter abbreviations in naming the contact-shuffled null models, where
each letter represents a retained correlation.  These correlations are (D)aily
pattern i.e. overall event frequency, (C)ommunity structure, (W)eight topology
correlation (B)ursty event dynamics on single links and (E)vent-event
correlations between links.  Below we reproduce verbatim the description of the
SBSW null models, more details about which can be found in the original paper:

\begin{itemize}
\renewcommand\labelitemi{--}
\item DCWB (\textit{equal-weight link-sequence shuffled}): Whole single-link
  event sequences are randomly exchanged between links having the same
  number of events. Temporal correlations between links are destroyed.
\item DCB (\textit{link-sequence shuffled}): Whole single-link event sequences
  are randomly exchanged between randomly chosen links.  Event-event
  and weight-topology correlations are destroyed.
\item DCW (\textit{time-shuffled}): Time stamps of the whole original event
  sequence are randomly reshuffled.  Temporal correlations are
  destroyed.
\item D (\textit{configuration model}): The original aggregated network is
  rewired according to the configuration model, where the degree
  distribution of the nodes and connectedness are maintained but the
  topology is uncorrelated.  Then, original single-link event
  sequences are randomly placed on the links, and time shuffling as
  above is performed.  All correlations except seasonalities like the
  daily cycle are destroyed.
\end{itemize}

\subsection*{Inducement Shuffling}
Our main results are based around the inducement-shuffled null models presented
below which are framed in terms of (T)imes, (L)ocations and (N)odes.  The input
to our inducement models are the session tuples.  The output is a set of new
session tuples of equal length with one or more of the correlations between
pairs of T, L and N destroyed.  We perform contact inference on the output
sessions using the method just described to arrive at a set of new contact
tuples.  We reiterate that because inducement shuffling results in a different
set of colocation events: it \emph{implicitly} induces a modified contact
network during contact inference, rather than explicitly reconfiguring the
contact network itself.

The listing below describes each of the new inducement null models.  Each
model's abbreviation is based on paired capital letters which indicate the
retained correlation(s) between times, locations and nodes.  For example, the
null model LN-TN retains (L)ocation/(N)ode and (T)ime/(N)ode correlations.  This
means that, compared to the original traces, individual nodes still visit the
same locations and they initiate sessions at the same start times. The change
from the original trace is shuffling individual sessions relative to locations,
thereby destroying the correlation between locations and times. Similarly, LN
retains only (L)ocation/(N)ode correlation.  Under this notation, the original
session trace could also be referred to as LN-TN-TL as correlations are retained
between all pairs of location, node and time.  Note that our inducement models
only destroy \emph{correlations} between the three variables---they do not alter
the independent frequency distribution of times, locations and nodes.  In other
words, nodes, locations and times with frequent activity maintain their activity
levels in the shuffled trace. Moreover, session \emph{durations} are always
retained.  That is, even if session times are shuffled, pairs of start and end
times always move together.  The inducement models are summarized in
Table~\ref{tbl:null_model_summary} in addition to the detailed listing below.
Note that the final column of Table~\ref{tbl:null_model_summary} includes
spreading prevalence at 1 day, the meaning of which will become clear when we
explain spreading dynamics shortly.

\begin{itemize}
\renewcommand\labelitemi{--}
\item LN-TN (\textit{maintains the location visitation and session start times
    of nodes, while destroying correlations between time and location}):
  sessions are first grouped by node.  For each grouping, the list of time pairs
  are shuffled (i.e. each start/end pair is exchanged with another pair in the
  same group).  Grouping by node acts to ``loosely bind'' nodes and
  times---although the times are shuffled, they are always reallocated to the
  same node and therefore only the binding between times and locations is
  destroyed.  i.e. LN and TN correlations are retained while TL correlation is
  destroyed.
\item TL-LN \textit{maintains the location visitation sequence of nodes and the
    correlation between time and location, while shuffling the node relation
    with session start time}: sessions are first grouped by location.  For each
  grouping, the list of nodes are shuffled.  Again, grouping acts as a loose
  binding mechanism keeping the same location/node association.  i.e. TL and LN
  correlations are retained while TN correlation is destroyed.
\item LN \textit{maintains the location visitation sequence of nodes, while
    destroying the correlation between session times and both locations and
    nodes}: time pairs of the entire trace are shuffled.  LN correlation is
  retained while TN and TL correlations are destroyed.
\item TN \textit{maintains the session times for nodes, while destroying the
    correlations between locations and both nodes and session times}: locations
  of the entire trace are shuffled.  TN correlation is retained while LN and TL
  correlations are destroyed.
\item TL \textit{maintains the correlation between session times and locations,
    while destroying correlations between nodes and both session times and
    locations}: nodes of the entire trace are shuffled.  TL correlation is
  retained while LN and TN correlations are destroyed.
\item The empty set $\emptyset$ \textit{destroys all pairwise correlations
    between time, location, and nodes, while maintaining the independent
    frequency distributions of all three variables}: locations of the entire
  trace are first shuffled and then nodes of the entire trace are shuffled.  All
  correlations between time, location and node are destroyed.
\end{itemize}

In total there are $2^{3} = 8$ potential models (including Original) based on
which if any of the correlation pairs chosen from (LN, TN, TL) are retained
during shuffling; however we omit one null model TL-TN from our analysis.  TL-TN
would in theory group by time and shuffle locations in order to destroy only the
location/node (LN) correlation.  Unlike location and node however, time is not a
categorical variable and so grouping would require a discretization of times
into an arbitrary number of slots.  It is not readily apparent why one number of
slots should be chosen over any other and so we leave contemplating TL-TN for
future work.

We note that in our trace, locations are inherently discrete---individual APs
with known identifiers.  The inducement-shuffled null models do however
generalize to broader settings, provided locations can be discretized.  For
example, a trace of mobile wireless bluetooth contacts annotated with known GPS
coordinates might be discretized into a grid of lat/lon squares or
geographically clustered.  On the other hand, a contact network with no known
locations (e.g. email) is not amenable to shuffling with the models presented in
this paper.  Moreover, though we focus our own simulations on a contact trace
predicated on colocated wireless devices, the theme of times/locations/nodes is
generalizable in the context of contact inference.

\subsection*{Spreading Dynamics}
We model spreading atop of all contact networks using the Susceptible-Infected
(SI) infection model~\cite{infectious_diseases}.  Nodes in the SI model are in
one of two states: (S)usceptible or (I)nfected.  State change is unidirectional
with nodes graduating from S to I upon satisfying the infection condition---in
our trace, colocation of a susceptible and infected device.  Under the SI model,
infection prevalence is monotonically nondecreasing as a function of time
growing until all devices reachable from initial conditions are infected.

Similar to the work in ~\cite{SBSW}, we start by first infecting a randomly
chosen node at a randomly chosen contact event.  The chosen event's timestamp is
interpreted as the simulation trial start time $t = 0$.  We restrict the random
sampling of the initial event to the first 4 days of trace to ensure a minimum
of 10 days simulation ``runway'' before reaching the end of the trace.  This
4/10 partition provides ample sampling opportunity across peak and trough
traffic periods as well as weekdays and weekends while allowing enough
simulation runway to observe the prevalence growth pattern.  After sampling the
initial infection event, we discard those contacts that occur either prior to
the event or $>10$ days after the event.  This ensures all simulations run
exactly 10 days, preventing a non-uniform number of samples beyond the 10-day
period (e.g. only initial infection events sampled from within the first 12
hours of trace would have known prevalence at 13.5 days).  We proceed to
calculate the prevalence denominator $N$ as equal to the size of the Largest
Connected Component (LCC) of the aggregated 10-day contact network which
typically consists of the majority of all devices ($>95\%$) in the period.  Note
that on rare occasions the initially sampled device will fall outside of the
LCC.  In this case we simply resample until the device is a part of the LCC.
Starting at the initial infection and stepping chronologically through the
contact sequence we then simulate the SI model of ideal diffusion whereby a
susceptible device becomes infected upon contact with an already infected
device.  We denote the set of infected devices at time $t$ as $I(t)$ and the
infection prevalence at time $t$ as $P(t) = |I(t)|/N$.  We perform a total of
\trials{} random spreading trials (different starting nodes) and average the
results.  In the supporting material we also provide a comprehensive pairwise
comparison of spreading potential for all inducement-shuffled null models along
with standard errors around the averaged results.

\section*{Results and Discussion}
\subsection*{Spreading Under Contact Shuffling}
In Figure~\ref{fig:sbsw_prev_vs_time_lcc_true} we present the spreading results
under the pre-existing SBSW contact-shuffled null models on the St Lucia contact
sequence. The pronounced ``wavestep'' pattern here captures the daily cycle of
human activity well. We also note that the time to near full prevalence in our
trace is on the order of a few days, remarkably shorter than the order
($\sim 100$ days) in MCN~\cite{SBSW}. This is due to the fact that, in a Wi-Fi
network successive transmissions can occur within a short period from one
infected node to multiple susceptible nodes that simultaneously connect to the
same AP, accelerating the whole spreading process.

Interestingly, our results show that the diffusion rate in DCWB is substantially
faster than in Original, whereas DCWB and Original have been found to have a
nearly identical diffusion rate in MCN~\cite{SBSW}. This suggests that the
temporal correlation between contact links is the most substantial impediment to
the spreading in the Wi-Fi network. Further destroying link weights (DCB) or
bursty single-link event sequences (DCW) only leads to a slightly faster
diffusion. This also significantly differs from the observed effect of the
similar null models in MCN, where the diffusion rates follow DCW $>$ DCB $>$
DCWB.  We conjecture that link weights and bursty single-link event sequences
make little difference in our trace due to the relative homogeneity of link
weights---most device contacts occur only once (Table
\ref{tbl:repeat_contacts}).  This means destroying weight or bursty event
correlations does little to perturb our contact network.  By comparison, the MCN
in~\cite{SBSW} we calculate to have average link weight
$\overline{\chi} \approx 34$ ($306 \times 10^6$ calls on only $9 \times 10^6$
links), whereby destruction of link weights or bursty event sequences would lead
to substantial perturbation.  Finally, consistent with~\cite{SBSW} we find D to
spread fastest.  We note however that the marginal difference is less than
observed in the prior work.

\subsection*{Spreading Under Inducement Shuffling}
We now present our main result in
Figure~\ref{fig:ssr_prev_vs_time_lcc_true}---the subsequent rate of spreading as
a function of time after inducement shuffling of the session records, which are
averaged over $250$ trials. The histograms of the prevalence after one day in
different trials under each null model are shown in
Figure~\ref{fig:one_day_prevs_hists} with the characteristics summarized in
Table~\ref{tbl:null_model_summary}.

The original trace is found to spread slower than all of the inducement-shuffled
null models, indicating the combined effects of location/node, time/node and
time/location correlations result in the slowest spreading.  In other words, the
intrinsic coupling of movement patterns of the nodes in time and space seem to
be impeding spreading by limiting randomness of contacts. Destroying the
time/location correlation with LN-TN only marginally accelerates spreading and
further destroying the location/node correlation in TN produces little further
acceleration.  This suggests that location-specific movement dynamics, which may
attract nodes to some locations and repel them from other, are not the primary
impediment to spreading. In contrast, destroying the time/node correlation with
TL-LN substantially accelerates spreading.  Maintaining the location/node
correlations while destroying all temporal correlations (LN) also leads to
relatively quick spreading, which provides further support that time/node
correlations are responsible for slow spreading.  Though it is interesting that
some crossover is observed around 2--5 days between TL-LN and LN, there is
little statistical strength to the statement that LN is ever truly faster which
is made clear by comparing TL-LN and LN in
Figure~\ref{fig:ssr_prev_delta_sem_pairs_lcc_true_abs_lim} (Supporting
Information). TL and $\emptyset$ occupy the top echelon with respect to speed of
spreading.  This indicates two things: (i) location/node correlations have some
impeding effect on spreading (TL being faster than TL-LN) and (ii) although TL
produces more contacts than $\emptyset$ by keeping the correlation between time
and location, both have a similarly fast spreading. The underlying reasons for
this similarity will be explored in the next section.  In summary, time/node
correlation is found to be the main impediment to spreading with location/node
correlation playing a smaller role.

\subsection*{Contact Frequency and Spreading}

To further explore the observed spreading phenomena, we study the contact
frequency and its relation with the infection prevalence in each
inducement-shuffling null model. In the spreading process under each null model,
the infection prevalence at a specific time and the cumulative contact frequency
upon that time are correlated, as they are both a monotonically increasing
function of time. A question remains whether the infection prevalence is
uniquely determined by the contact frequency across different null models. In
other words, does infection prevalence and contact frequency have a direct
causality. As we show in the following, there is no simple causality between
these two, suggesting that the contact frequency is not sufficient to
distinguish the spreading potential in different null models.

\subsubsection*{Contact Characterisation}
We begin by characterising the contact and session statistics over time and
space in Figures \ref{fig:active_sessions} through
\ref{fig:contacts_per_node_ecdf_uniq_lcc_true}.
Figure~\ref{fig:active_sessions} shows the number of active sessions over time,
clearly showing the differences in session volumes over weekdays and weekends
and the diurnal patterns.  We expect this to result in higher spreading
propensity during weekdays when there are significantly more
contacts. Figure~\ref{fig:locations_per_node_ecdf_lcc_true} shows the ECDF of
the unique locations visited per node for each of the inducement null models. A
clear clustering of 2 sets of null models emerges, where all the models that
destroy the location-node correlation (omit LN from their name) result in a
higher number of unique locations visited by each node.  Figures
~\ref{fig:contacts_per_node_ecdf_total_lcc_true} and
~\ref{fig:contacts_per_node_ecdf_uniq_lcc_true} compare the total and unique
contacts per node, or put another way, the node degrees in the networks based on
contacts. The total contacts distribution is comparable for all null models,
with slight differences among the models. For unique contacts, we observe more
pronounced differences between the models, where models that destroy LN (TL, TN,
and $\emptyset$), produce the largest number of unique contact pairs. This
effect is likely a result of the greater number of unique locations produced by
these null models, which increases the likelihood of new unique contacts. LN
produces the second highest number of unique contacts, while preserving any
temporal correlation with LN results in fewer yet similar results. The original
trace produces the least contacts, confirming that the inherent spatiotemporal
node correlations limits unique contact opportunities.

The null models also impact the intersession times for each node, as shown in
Figure~\ref{fig:intersession_time_ecdf_lcc_true}. A clear separation of two
groups of null models emerges between null models that break the node relation
with session start time (omit TN), and those that preserve it. All null models
that preserve TN have shorter intersession time, as the the temporal correlation
between successive sessions is apparent. Large spikes in the plot account for
typical session renewal times, such as at around 1 second when multiple attempts
to establish a session occur, or at 5, 10, 15, and 20, 25, and 30 minutes, where
access points typically update session status every 5 minutes. For null models
that break TN, intersession times are significantly longer, emphasising the
contribution of the time-node correlation to shorter intersession times.

\subsubsection*{Cumulative Contact Volume}
We turn our attention next to elucidating the total contact counts in
Figure~\ref{fig:ssr_contact_count_vs_time_total}. This effectively captures the
total number of links in the network over time based on observed contacts, in
contrast with the degree distribution shown above that illustrated comparable
degree distributions for all null models.  All contact tallys exhibit the
characteristic diurnal stepping pattern of five weekdays interleaved with two
weekend days.  Original and TL grow at essentially the same rate (overlapping
lines) as they both retain the correlation between times and locations leading
to the same number of total colocation events.  We say ``essentially'' rather
than ``exactly'', as all null models which decorrelate time and node (TL-LN, LN,
TL, $\emptyset$) can occasionally produce ``imaginary'' contact events whereby a
node is colocated with itself.  Such events are discarded during contact
inference and so are not accounted for in our tallys.  For TL we find these
imaginary contacts account for a negligible portion of all contact events ($<$
0.03\%) and so TL appears to overlap Original.  TL-LN on the other hand clearly
illustrates the imaginary event phenomenon.  In fact, the delta between Original
and TL-LN in Figure ~\ref{fig:ssr_contact_count_vs_time_total} is a direct
measure of the the number of discarded imaginary contacts under TL-LN shuffling
which we find typically runs around 3\%.  It is interesting to note that LN is
the only other shuffling with non-negligible imaginary events, again around 3\%.
The implication is that individual nodes likely have a strong affinity for
specific locations (retained by *LN) leading to a (relatively) high number of
imaginary events when the timing of node's presence in the network is shuffled.

LN-TN produces the fewest total contacts indicating that correlated times and
locations are the largest driver of contact volume in the Original trace.
Further destroying either the location/node or time/node correlation as is done
by TN and LN respectively increases total contacts.  We expect that TN, LN and
$\emptyset$ all produce the same number of total contacts as they all have the
same expected number of nodes at a given location at a given time.  The observed
disparity of LN having fewer total contacts than TN and $\emptyset$ is simply
attributable to the $\approx$3\% of imaginary contacts produced and discarded
under LN shuffling.

We now focus to the unique contact counts in
Figure~\ref{fig:ssr_contact_count_vs_time_unique}.  Again, all contact tallys
exhibit the characteristic diurnal stepping pattern being strongly driven by
macro-scale periodic activity in the network.  Whereas the original trace
produces the most total contacts (in a tie with TL and TL-LN before discarding
imaginary contacts), it also produces the \emph{least} unique contacts.  This
suggests the ``mixing'' effect of the null models which increases unique
contacts tends to be stronger than the opposite hindering effect the null models
sometimes have on total contact volume.  LN-TN and TL-LN produce approximately
the same number of unique contacts which implies that destroying either
time/location or time/node correlation has about equal effect on unique
contacts.  Destroying time/location and time/node correlations leaving only
location/node correlation (LN) further increases unique contact.  TN and
$\emptyset$ produce still more unique contacts, possibly because the
aforementioned affinity of devices for specific locations which is retained in
LN had a stronger impeding effect on new unique contacts than does node's time
preferences.  TL leads to the most unique contacts, likely because it retains
the total contact boost driven by node's preferences to be in the same location
at the same time with the added benefit of mixing which nodes partake in the
contact event.

As a final point of comparison between Figures
~\ref{fig:ssr_contact_count_vs_time_total} and
~\ref{fig:ssr_contact_count_vs_time_unique} we note how unique contacts are
always within a factor of 2.5 of total contacts.  Again referring to
Table~\ref{tbl:repeat_contacts}, this is likely a by-product of a contact
network whose original link ``weights'' as measured by repeat contact count are
quite low versus the MCN analyzed in~\cite{SBSW} (average link weight
$\overline{\chi} \approx 34$).

Already Figures ~\ref{fig:ssr_contact_count_vs_time_total} and
~\ref{fig:ssr_contact_count_vs_time_unique} appear to suggest that there is a
weak relationship between a null model's propensity to produce contact events
and its subsequent spreading potential.  We now proceed to plot contacts versus
prevalence where this will become even more obvious.

\subsubsection*{Contacts vs. Prevalence}
Figure~\ref{fig:ssr_prev_vs_contacts_total_lcc_true} and
Figure~\ref{fig:ssr_prev_vs_contacts_unique_lcc_true} plot the infection
prevalence as a function of the number of total and unique contacts
respectively.  It is evident that the relation between the infection prevalence
and the contact frequency varies in different null models, suggesting that no
direct causality exists. That is to say, the time of reaching a given prevalence
in each null model is different, and the number of contact events upon this time
is different as well. Therefore, without knowing the type of the null model, one
is not able to infer the infection prevalence directly from the contact
frequency. We can also interpret this phenomenon by using the term ``spreading
efficiency'', i.e. the number of contacts required to reach a specific
prevalence. Our results demonstrate that the proposed null models have different
spreading efficiency.

It is interesting to note that this disparity in spreading efficiency has also
been implicitly observed both in this paper and prior work~\cite{SBSW} under
contact shuffling.  This follows from the fact that contact shuffling inherently
retains the same total contact frequency as a function of time, even after
shuffling.  Although unique (non-repeat) contacts may deviate from the original
trace throughout the simulation period under D and DCW shufflings, these also
equalize to that of the original trace by the end of the simulation.  In any
case, DCWB and DCB which match Original in both total and unique contacts at
each time step present with drastically superior diffusion potential (on a
per-contact basis) than the original trace in our analysis.  We suggest that
future work may wish to further explore the relationship between both the static
and temporal structures of contact-shuffled and inducement-shuffled networks.
Doing so may reveal common emergent properties such as statistical similiarities
that offer a unified explanation to the disparity of spreading efficiency.

\section*{Conclusion}
This paper has analyzed impediments and catalysts to spreading in a contact
network through the introduction of a set of inducement-shuffled null models
which separately destroy the correlations between times, locations and nodes.
The inducement-shuffled null models have enabled second-order causal reasoning
about the observed spreading propensity.  That is, how is spreading affected by
the idiosyncratic behavior that lead to the observed contact network in the
first place, rather than how the observed contact network alone affects
spreading.  Among our main observations is that (i) spreading is primarily
impeded by time/node correlation and (ii) though correlations have a slowing
effect in general, retaining time/location correlation alone proves an exception
which slightly increases the rate of spreading versus a network with all
correlations destroyed.  Furthermore we have demonstrated a curious disparity
between a null model's ability to produce frequent contact events and its
propensity to promote spreading.  Finally, we have found that under
pre-existing contact-shuffled null models temporal link correlations are the
main spreading impediment in our trace, in contrast to earlier reported
results.  

We conclude by proposing a number of avenues for future work.  Firstly, there
are several areas already alluded to in this manuscript.  These are (i) defining
and justifying appropriate grouped shuffling models on non-categorical variables
such as time (ii) establishing if and how ``imaginary'' contact events between a
node and itself ought to ever be avoided in shuffling and (iii) clearly
articulating the conditions which may lead to large disparities between contact
volume and prevalence.  More broadly, it would be useful to explore whether the
abstraction of times, locations and nodes generalizes well to other data traces
of colocation contact networks, and indeed to other types of contact networks,
particularly those which do not operate at human scale (e.g. cellular level
networks).  Complex networks in itself offers a unified framework for
contemplating a diverse range of systems and so we expect that some subset of
these may well be framed in a similar null model construction to that used in
this paper.  Moreover, there may exist alternative null model abstractions that
will afford the same second-order causal reasoning about the nature of spreading
in other types of contact networks.  We urge the reader to consider what these
abstractions might be.


\section*{Figure Legends}
\figSBSWPrevVsTimeLCCTrue{}
\figSSRPrevVsTimeLCCTrue
\figOneDayPrevsHists
\figActiveSessions
\figLocationsPerNodeECDFLCCTrue
\figContactsPerNodeECDFTotalLCCTrue
\figContactsPerNodeECDFUniqLCCTrue
\figIntersessionTimeECDFLCCTrue
\figSSRContactCountVsTimeTotal
\figSSRContactCountVsTimeUnique
\figSSRPrevVsContactsTotalLCCTrue
\figSSRPrevVsContactsUniqueLCCTrue

\section*{Tables}
\makeatletter{}%
\begin{table}[!h]
\caption{Null model summary and prevalence at 1 day $\pm$ standard error of the mean.}
\label{tbl:null_model_summary}
\centering
\begin{tabular}{l c c c r}
\hline
Shuffling & LN & TN & TL & $|I(1\:day)|/N$\\
\hline
Original & $\checkmark$ & $\checkmark$ & $\checkmark$ & 33.4\%$\pm$0.6\%\\
Group node, shuff. time & $\checkmark$ & $\checkmark$ & & 35.0\%$\pm$0.6\%\\
Group location, shuff. node & $\checkmark$ & & $\checkmark$ & 44.6\%$\pm$0.8\%\\
Shuffle time & $\checkmark$ & & & 46.5\%$\pm$0.8\%\\
Shuffle location & & $\checkmark$ & & 37.3\%$\pm$0.6\%\\
Shuffle node & & & $\checkmark$ & 50.5\%$\pm$0.8\%\\
Shuffle location, shuffle node & - & - & - & 50.3\%$\pm$0.7\%\\
\hline
\end{tabular}
\end{table}
\makeatletter{}%
\begin{table}[!h]
\caption{Number of repeat contacts between pairs of nodes.}
\label{tbl:repeat_contacts}
\centering
\begin{tabular}{r r}
\hline
repeats & count\\
\hline
1 & 385 453\\
2 & 70 092\\
$>$ 2 & 96 298\\
\hline
\end{tabular}
\end{table}

\section*{Supporting Information Captions}
We include Figures~\ref{fig:ssr_prev_delta_sem_pairs_lcc_true} and
~\ref{fig:ssr_prev_delta_sem_pairs_lcc_true_abs_lim} to enable the interested
reader to more closely compare pairwise prevalences under the
inducement-shuffled null models.
Figure~\ref{fig:ssr_prev_delta_sem_pairs_lcc_true} is designed primarily for
macro-scale comparison of prevalences, i.e. approximately how much faster or
slower is one shuffle versus another.  By comparing specific row and columns the
reader can easily see any large difference in prevalences as a function of time
of a specific pair of shufflings as the y-axis scale encompasses the maximum
absolute differences.
Figure~\ref{fig:ssr_prev_delta_sem_pairs_lcc_true_abs_lim} on the other hand is
designed for micro-scale comparison of prevalences.  The y-axis has been zoomed,
which hides differences of large magnitude but allows the reader to more clearly
visualize standard errors and overlaps between prevalence pairs.  This is
designed in particular to allow the reader to gauge whether small but
perceivable prevalence differences in Figure~\ref{fig:ssr_prev_vs_time_lcc_true}
are of much statistical significance.  For example, the reader might notice the
brief crossover between LN and TL-LN in
Figure~\ref{fig:ssr_prev_vs_time_lcc_true} and be curious how significant this
result is.  By comparing LN and TL-LN in
Figure~\ref{fig:ssr_prev_delta_sem_pairs_lcc_true_abs_lim} one is then able to
ascertain that there exists relatively large standard errors around both LN and
TL-LN at the time of this measurement and so it is not likely to hold that TL-LN
is ever faster than LN with much statistical significance.  On the other hand,
the reader may be interested in where there are clear differences, such as
between TL and LN where there is a substantial amount of space between the two
prevalences (TL being faster).

\figSSRPrevDeltaSEMPairsLCCTrue

\figSSRPrevDeltaSEMPairsLCCTrueAbsLim

\end{document}